# Conventional superconductivity at 190 K at high pressures

A.P. Drozdov, M. I. Eremets*, I. A. Troyan

*Max-Planck Institut fur Chemie, Chemistry and Physics at High Pressures Group*

*Postfach 3060, 55020 Mainz, Germany*

*e-mail: m.eremets@mpic.de

The highest critical temperature of superconductivity $T_c$ has been achieved in cuprates[1]: 133 K[2] at ambient pressure and 164 K at high pressures[3]. As the nature of superconductivity in these materials is still not disclosed, the prospects for a higher $T_c$ are not clear. In contrast the Bardeen–Cooper–Schrieffer (BCS) theory gives a clear guide for achieving high $T_c$: it should be a favorable combination of high frequency phonons, strong coupling between electrons and phonons, and high density of states. These conditions can be fulfilled for metallic hydrogen and covalent hydrogen dominant compounds[4,5]. Numerous followed calculations supported this idea[6,7] and predicted $T_c$=100-235 K for many hydrides[6] but only moderate $T_c$~17 K has been observed experimentally[8]. Here we found that sulfur hydride transforms at P~90 GPa to metal and superconductor with $T_c$ increasing with pressure to 150 K at ≈200 GPa. This is in general agreement with recent calculations of $T_c$~80 K for $H_2S$ [7]. Moreover we found superconductivity with $T_c$≈190 K in a $H_2S$ sample pressurized to P>~150 GPa at T>220 K. This superconductivity likely associates with the dissociation of $H_2S$, and formation of $SH_n$ (n>2) hydrides. We proved occurrence of superconductivity by the drop of the resistivity at least 50 times lower than the copper resistivity, the decrease of $T_c$ with magnetic field, and the strong isotope shift of $T_c$ in $D_2S$ which evidences a major role of phonons in the superconductivity. $H_2S$ is a substance with a moderate content of hydrogen therefore high $T_c$ can be expected in a wide range of hydrogen–contain materials. Hydrogen atoms seem to be essential to provide the high frequency modes in the phonon spectrum and the strong electron-phonon coupling.

A room temperature superconductor probably is the most desired system in solid state physics[10]. So far the greatest advances, cuprates[1], pnictides[11] and number of others were obtained in a serendipitous way. As there is no clear theory for these superconductors, it is difficult to predict where progress will be made. Another, distinct way is the BCS theory described by the pairing of electrons mediated by phonon lattice vibrations. The Eliashberg formulation of this theory puts no apparent bounds on $T_c$[12]. In practice, however BCS theory does not have a predictive power to calculate material-specific properties[12] but it serves as a guide for the search for superconductors. Materials with light elements are especially favorable as they provide high frequencies in the phonon spectrum. Indeed many superconductive materials have been found in this way, but with a disappointing moderately high $T_c$ of about 10 K. The highest $T_c$=39 K that has been found in this search is in $MgB_2$. There is still frequent preconception that $T_c$~30 K is the maximum what can be achieved in conventional superconductors.

N. Ashcroft[4] turned attention to hydrogen as the lightest element for which high $T_c$ can be expected, because of the very high vibrational frequencies due to the light hydrogen atom, a strong electron-phonon interaction, and covalent bonding. Further calculations showed that metallic hydrogen should be a superconductor with a very high critical temperature $T_c$ ~100-240 K[13-15] for molecular hydrogen, and $T_c$ = 300-350 K in the atomic phase at 500 GPa[16]. However superconductivity in pure hydrogen has not yet been found while the conductive and likely semimetallic state of hydrogen has been recently achieved[17]. Hydrogen dominate materials such as covalent hydrides $SiH_4$, $SnH_4$ etc. might also be good candidates for high $T_c$ superconductivity[5]. Similar to pure hydrogen, they have a high Debye temperature. Moreover, heavier elements might be beneficial as they contribute to the low frequencies that enhance electron phonon coupling. Lower pressures are required to metallize these hydrides in comparison to pure hydrogen. Ashcroft's general idea was supported in numerous calculations for $T_c$ of different hydrogen containing materials. The key in this new approach of searching for new superconductors is the prediction of structures from *ab-initio* calculations[6,18,19]. As soon as the structure is known, the electronic phonon spectra can be calculated and $T_c$ estimated by solving the Eliashberg equations. However, the theoretical search for structures at present cannot guarantee the most stable structure. Regularly new theoretical results are published where the predicted structures are revised and lower energy structures are found[6,7]. It can only be determined experimentally if the calculated $T_c$s are correct and give prospects for high $T_c$ in the hydrogen dominate materials. However, experimental progress is much slower[8,20] apparently because the majority of the theoretically studied materials are difficult to access, and high pressure studies are difficult and challenging, especially electrical measurements. For the present study we selected $H_2S$ because it is relatively easy to handle, transforms to metal at a low pressure of ≈100 GPa, and the calculated $T_c$=80 K[7] is high.

$H_2S$ is known as a typical molecule substance with a rich phase diagram[21]. At about 96 GPa hydrogen sulfide transforms to metal[22]. The transformation is complicated by the partial dissociation of $H_2S$ and the appearance of elemental sulfur at P>27 GPa at room temperature, and higher pressures for lower temperatures[21]. Therefore, the metallization of hydrogen sulfide can be explained by elemental sulfur which is known to become metallic above 95 GPa[23]. No experimental studies on hydrogen sulfide are known above 100 GPa. Recent theoretical work[7] revised the phase diagram of $H_2S$, and a number of new stable structures were found. At P>130 GPa – higher pressures than experimentally observed – hydrogen sulfide was predicted to become a metal and a superconductor with a maximal transition temperature of ~80 K at 160 GPa. Precipitation of sulfur has been found to be very unlikely, in apparent contradiction to the experiments[21,24].

In our typical experiments the Raman spectra for $H_2S$ and $D_2S$ measured during the pressure increase are in general agreement with the literature data[25,26] (see Extended Data Fig. 1). In electrical measurements, $H_2S$ starts to conduct at P~50 GPa. At this pressure it is a semiconductor as follows from the temperature dependence of resistance and a pronounced photoconductivity. At 90-100 GPa resistance further drops, and the temperature dependence becomes metallic. No photoconductive response is observed in this state, or the resistance increases under illumination. $H_2S$ is a poor metal – its resistivity at ~100 K is $\rho$≈3 $10^{-5}$ Ohm m at 110 GPa and $\rho$≈3 $10^{-7}$ Ohm m at ~200 GPa.

During cooling of the metal phase to 4K at pressures of about 100 GPa (Fig. 1a) resistance abruptly drops three to four orders of magnitude at $T_c \approx 23$ K. After warming to T~100 K, pressure was increased to the next value and the sample was cooled. The next loadings were performed at 100-150 K (Fig. 1a, see Extended Data Fig. 2). $T_c$ increased with pressure (Fig. 2a) with sharp growth when pressure approached 200 GPa. We found a route to another superconductive state with $T_c \approx 190$ K by application of pressure P>~150 GPa but at significantly higher temperatures of 220-300 K (Fig. 2b). This $T_c$ has weak pressure dependence (Fig. 2b) – very different from the $T_c$ obtained at low temperatures (Fig. 2a). Frequently the 190 K step is accompanied with a step with $T_c$~30 K which disappears with time (>~ 1 day) or further application of pressure while the 190 K step sharpens (see Extended Data Fig.3). There are some oscillations on the R(T) with period of 25-30 K clearly seen in a number of runs (see Extended Data Fig. 3). The accumulated data reveal a quite complex P-$T_c$ diagram and we separate it for two parts: Fig. 2a and Fig.2b according to the different routes of achieving $T_c$s.

We conclude that for both the routes the observed steps at the temperature dependence of resistance (Figs 1, 3.4) represent a transition to the superconducting state for the following arguments:

(1) The measured minimum resistance is at least as low $\approx 10^{-11}$ Ohm m – about two orders of magnitude less than for pure copper (Figs 1, see Extended Data Figs. 2,3) measured at the same temperature[27].

(2) $T_c$ shift to lower temperatures with the available magnetic field up to 7 Tesla (Fig.3). Much higher fields are required to destroy the superconductivity: extrapolation of $T_c(B)$ gives an estimation of critical magnetic field at ~ 25 T and 70 T for two critical temperatures (Fig. 3).

(3) A strong isotopic effect: $T_c$ shifts to lower temperatures for $D_2S$ (Figs 1,2, see Extended Data Fig. 4). For the phonon mechanism of the pairing of electrons in the BCS superconductivity the critical temperature depends on the square root of the mass: $T_c \sim M^{-\alpha}$ where $\alpha_H \approx 0.5$. For both superconducting states of sulfur hydride with $T_c \approx 60$ K and $T_c \approx 185$ K (Fig 2) we observe a critical temperature of the corresponding isotope at 30 K and 90 K indicating phonon assisted superconductivity.

The behavior of $T_c$ with pressure (Fig.2a) reasonably agrees with the calculations for $H_2S$[7]: the experimental values of $T_c$s are comparable with the predicted $T_c \approx 80$ K which is apparently conventional superconductivity. The pressure dependence is different, however, there is no predicted drop of $T_c$ at 160 GPa[7], or this drop might be at P >~200 GPa instead.

The superconductivity at $T_c$~190 K shown in the Fig. 2b is not seen or missed in the calculation for $H_2S$[7]. High temperatures of T>220 K are required to reach the $T_c$~190 K therefore decomposition of $H_2S$ can be involved. While precipitation of elemental sulfur can be expected, which is well known at low pressures of P>27 GPa[21], the superconductivity of sulfur alone is too low (Fig. 2a, see Extended Data Fig. 5). Precipitation of sulfur might be of particular interest as likely, sulfur forms impurities or clusters in a host lattice which can promote an increase of $T_c$ through interface effects, instabilities, and disproportionation which are common features of high temperature superconductors[3]. The sharp increase of $T_c$ at P>~200 GPa (Fig. 2a) can probably be explained by an increase of electron density of states due to impurities of sulfur. The effect of precipitation on superconductivity might be interesting to study in other systems too[3].

Another expected product of decomposition of $H_2S$ is hydrogen. However the strong characteristic vibron from $H_2$ molecule was never observed in the Raman spectra in spite of using a sensitive spectrometer and ultralow luminescence synthetic diamond anvils. No $H_2$ vibron has been found in Ref. [21] either even after 1 hour of accumulation. We suppose therefore that the dissociation of $H_2S$ might go differently: $2H_2S \rightarrow H_4S + S$, or $3H_2S \rightarrow H_6S + S$. It is natural to expect these reactions as the valency of sulfur is 2, 4, and 6. The molecules $H_4S$ and $H_6S$ are known to be thermodynamically unstable, but kinetically stable at ambient pressure[8], and high pressure might stabilize them. In fact calculations[7] indirectly support this hypothesis as the dissociation $H_2S \rightarrow H_2 + S$ was shown to be energetically very unfavorable. We found further theoretical support for our hypothesis on the dissociation of $H_2S$ to $H_nS$ (n>2) and sulfur in Ref [9]. In this work the van der Waals compound[28] ($H_2S)_2H_2$ has been considered and it has been shown that at pressures above 111 GPa $H_3S$ molecular unit is built, and that above 180 GPa, sulfur and hydrogen form a lattice with coordination number of the S atom is six. The predicted $T_c$s ~160 K and 190 K for these two phases are close to our experimental values and $T_c$ also decreases with pressure above 180 GPa (Fig. 2b). Moreover, our Raman measurements of the sample with $T_c$~190 K showed a phase transformation at 180 GPa (see Extended Data Fig. 5). Thus $T_c$~190 K (Fig. 2b) might be related to superconductivity of $H_3S$ obtained as a result of the dissociation of $H_2S$. Further theoretical and experimental studies of $H_2S$ and higher hydrides $H_nS$, n>2 are apparently needed.

We have found high $T_c$ superconductivity in $H_2S$ – material with low fraction of hydrogen. This probably will give a prospect to find high $T_c$ in other hydrogen contain materials first of all various carbon-based materials: fullerenes, aromatic hydrocarbon, graphanes etc. Instead of application of pressure they can be turned to superconducting state by doping or gating.

**Methods Summary**

A diamond anvil cell (DAC) was placed into a cryostat and cooled down to ≈200 K (within the temperature range of liquid $H_2S$) and then $H_2S$ gas was put through a capillary into a rim around diamond anvil where it liquidified. $H_2S$ of 99,5% and $D_2S$ of 97% purity have been used. Liquid $H_2S$ was clamped in a gasket hole. After the clamping, DAC was heated to ~220 K to evaporate the rest of the $H_2S$, and the pressure was further increased at this temperature. A metallic gasket of DAC was separated from the electrodes with an insulating layer made of Teflon or NaCl as these materials do not react with $H_2S$. Typical dimensions of the sample are a diameter ≈10 μm and a thickness of ≈1 μm. The material of electrodes was Ti covered by Au to protect Ti from oxidation. Four electrodes were sputtered on the anvil. To check a possible contribution of the diamond surface to conductivity we prepared a different configuration of electrodes: two electrodes were sputtered on one anvil and another two on another anvil similar to Ref. [17]. Electrical and Raman measurements were done in the same cryostat. Resistance was measured by using the four-probe Van der Pauw method (Fig. 1 and see

Extended Data Fig. 3) with a current of 10 -10000 µA. The dependence of superconducting transitions on the magnetic field has been measured with a small nonmagnetic DAC in a Quantum Design physical property measurement system (PPMS6000) in a 4-300 K temperature range up to 7 Tesla. Pressure was determined by a diamond edge scale[29]. For optical measurements a Raman spectrometer was equipped with a nitrogen-cooled CCD and notch filters. The 632.8 nm line of a He-Ne laser was used to excite the Raman spectra and to determine pressure.

Acknowledgments. Support provided by the European Research Council under the 2010-Advanced Grant 267777 is gratefully acknowledged. We appreciate support provided by Prof. U.Pöschl. The authors are thankful for V. Ksenofontov and B. Balke for generous help with measurements at PPMS and useful discussions.

Author contributions: A. D. performed the whole experiment and contributed to the data interpretation and writing the manuscript. M.E. designed the study, wrote the manuscript and participated in the experiments. I.T. participated in experiments and discussions. M.E., and A. D. contributed equally to this paper.

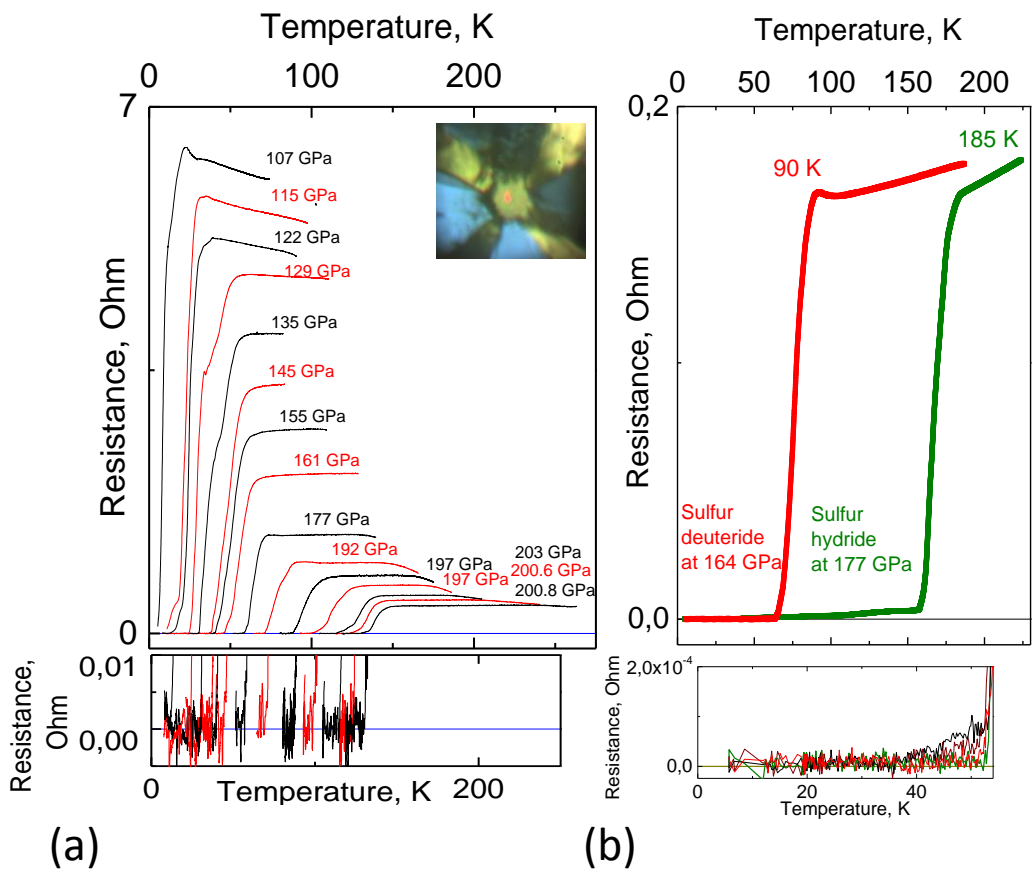

Fig. 1. Temperature dependence of resistance of sulfur hydride and sulfur deuteride measured at different pressures. The pressures did not change during the cooling (within ≈5 GPa). Resistance was measured with four electrodes deposited on a diamond anvil touched the sample (photo). Diameter of the samples was ~25 μm and the thickness ~1 μm.

(a) Sulfur hydride as measured at the growing pressures, the values are indicated near the corresponding plot. Plots at pressures <135 GPa were scaled (reduced in 5-10 times) for easier comparison with the higher pressure steps. The resistance was measured with current of 10 μA. Bottom: the resistance plots near zero. (b) Comparison of the superconducting steps of sulfur deuteride and hydride at similar pressures. Bottom: resistance measured near zero. Resistance was measured in four channels with van der Pauw method (SI Fig. 4) with current of 10 mA.

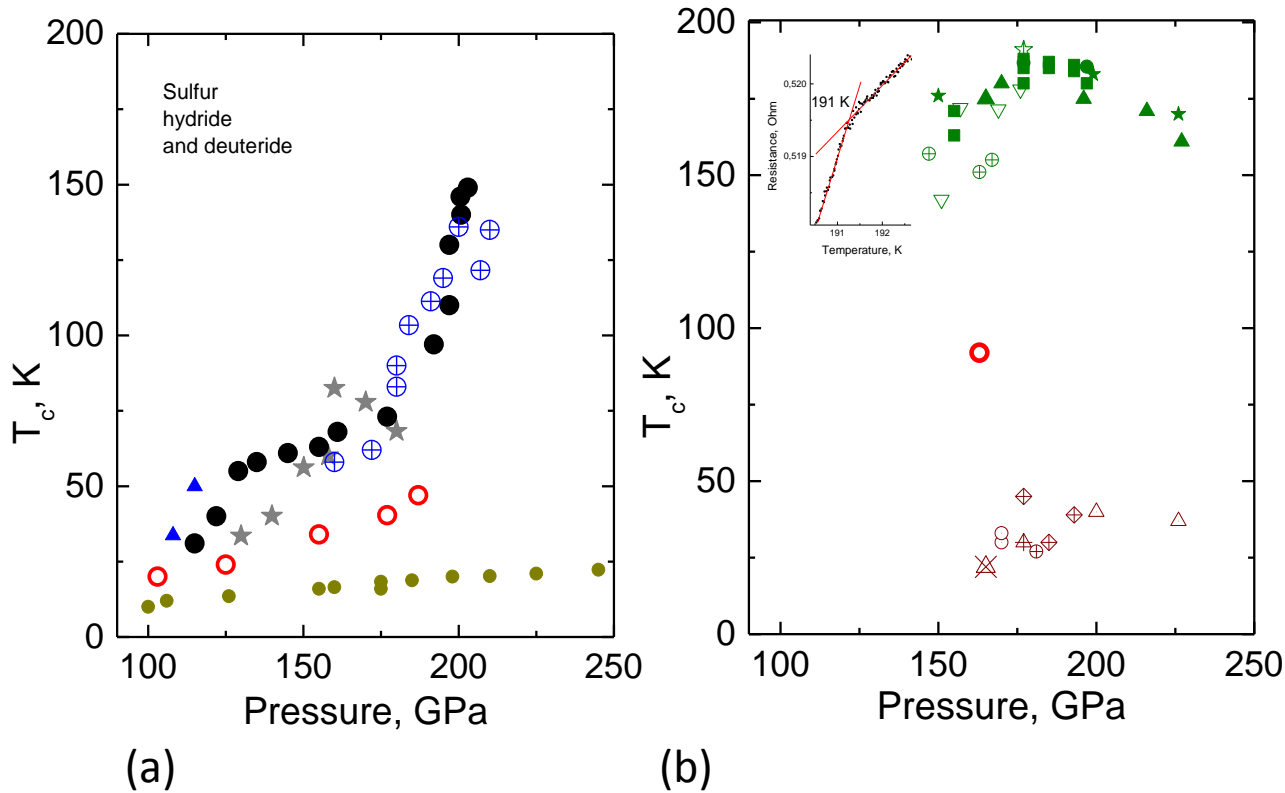

Fig. 2. Pressure dependence of critical superconducting temperature $T_c$ on pressure. Only four probe electrical measurements in van der Pauw geometry are presented in both panels. Critical superconducting temperature $T_c$ was determined as a point where resistance starts to sharply decrease with cooling from the plateau at the R(T) plot (inset in (b)).

(a) Data were obtained when pressure was applied in the 100-190 GPa pressure range at 100-150 K, and higher temperatures at P~200 GPa when $T_c$ sharply increased. Black points are data from Fig. 1a. Blue points - other runs. Red points are measurements of $D_2S$. Dark yellow points are $T_c$s of pure sulfur. Grey stars are calculations from Ref. 7.

(b) Higher $T_c$~190 K found when pressure P> 150 GPa was applied in combination with 220-300 K temperatures. The 190 K step is accompanied with another step with $T_c$~30 K (wine points). It is shown in SI Fig. 3. The red point is $T_c$ for $D_2S$ sample (Fig. 1b).

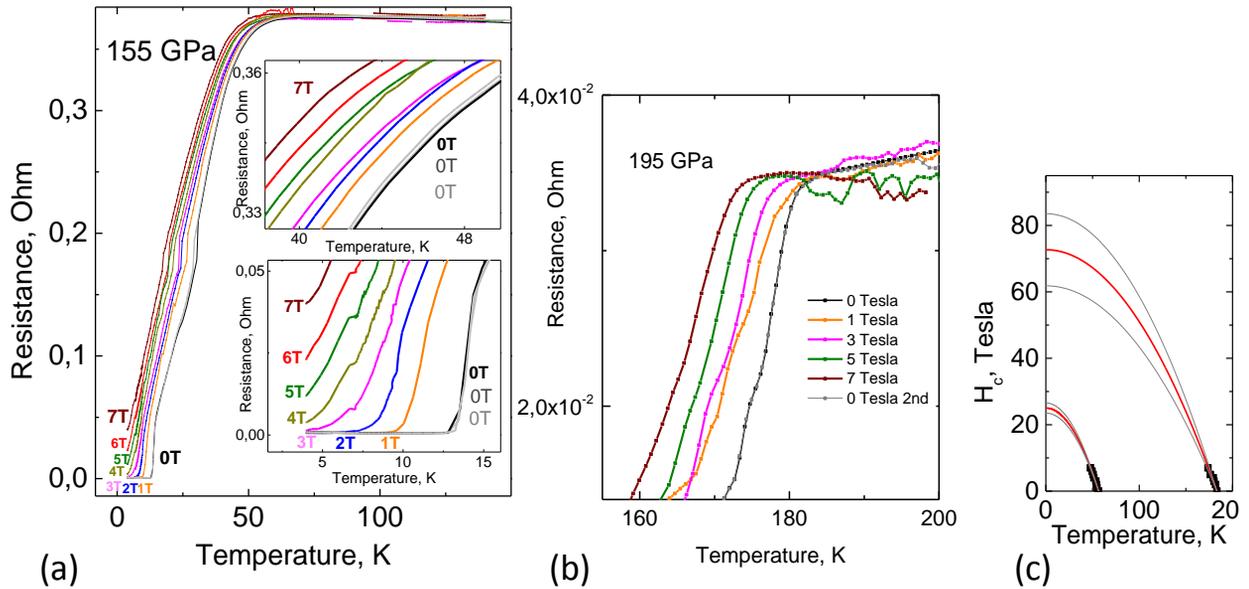

Fig. 3. Temperature dependence of resistance of sulfur hydride at different magnetic fields.
(a) The shift of the ~60 K superconducting step in the 0-7 T magnetic fields. The upper and low parts of the step are enlarged in the insets. Temperature dependence of resistance without magnetic field was measured three times: before applying a field, after applying 1,3,5,7 Teslas and at the end of measurements (black, grey and dark grey colors).
(b) The same measurements but with the 185 K step. (c) The shift of critical temperature of superconducting transition $T_c$ with magnetic field. The plots were extrapolated to high magnetic fields with $H_c(T)=H_{c0}(1-(T/T_c)^2)$ formula to estimate the critical magnetic field $H_c$. The extrapolation has been done with 95% confidence band (grey lines).

*Supplementary information*

# Conventional superconductivity at 190 K at high pressures

**A. P. Drozdov, M. I. Eremets\*, I. A. Troyan**

*Max-Planck Institut fur Chemie, Chemistry and Physics at High Pressures Group*
*Postfach 3060, 55020 Mainz, Germany*
*e-mail: m.eremets@mpic.de

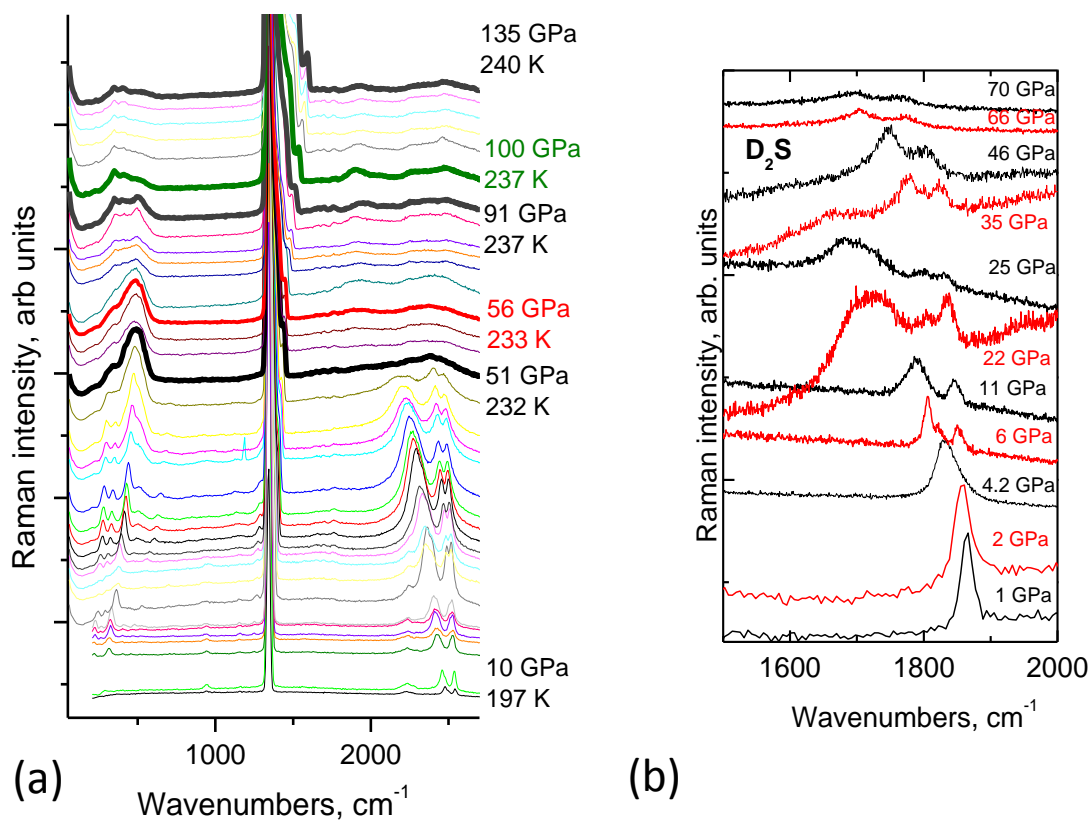

Fig. 1. Raman spectra of sulfur hydride at different pressures. (a) Spectra at increasing pressure at ~230 K. The spectra are shifted each other. The temperature of the measurement is also indicated. (b) Raman spectra of $D_2S$ measured at T~170 K.

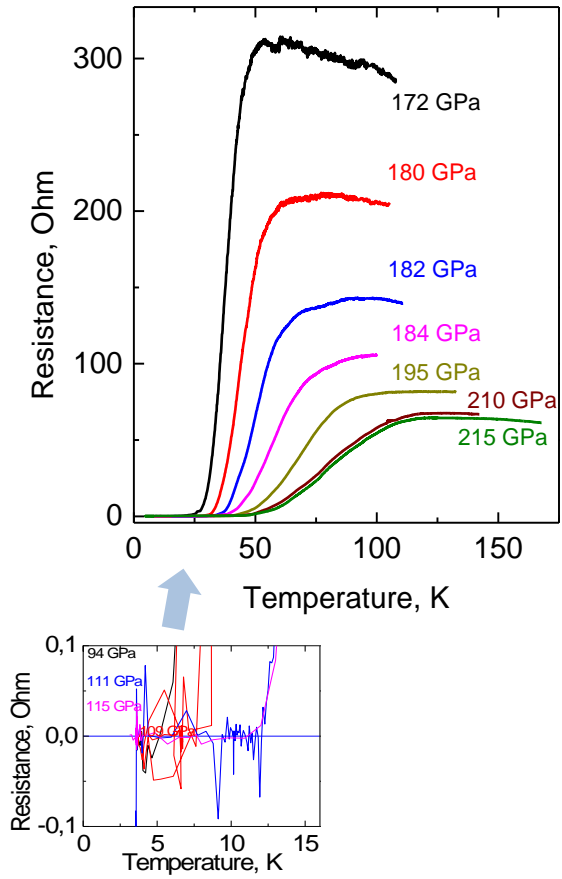

Fig. 2 . Temperature dependence of resistance (superconducting steps). Corresponding T$_c$s are shown by blue points in Fig. 2.  Left figure demonstrates minimum resistance measured in this run with current of 10 µA.

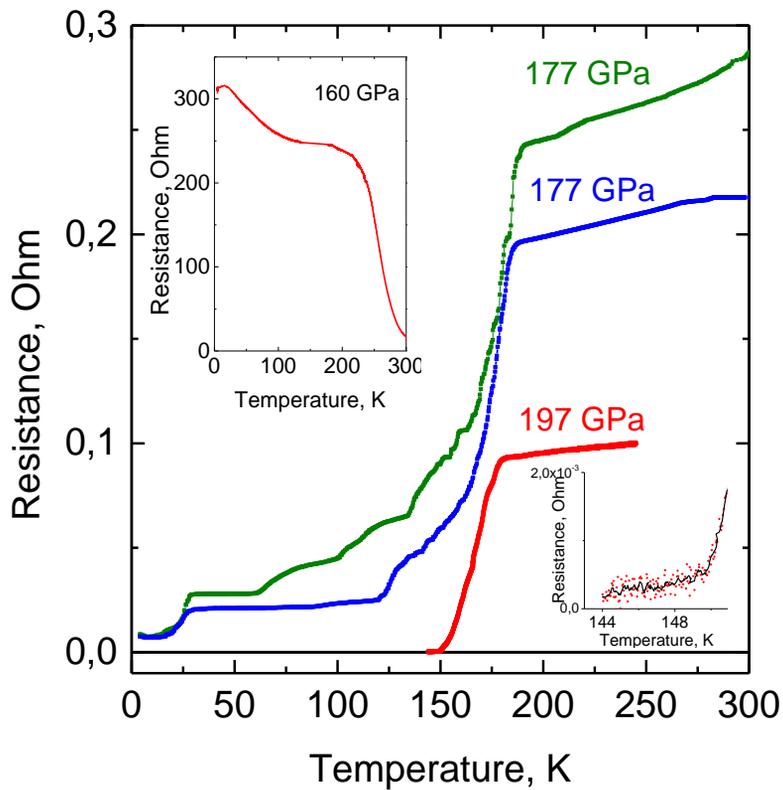

Fig. 3. Transformation of the superconducting state in the $H_2S$ sample with pressure, temperature and time.
At pressures up to 155 GPa there is only one SC step at ~60 K. After warming to 300 K at this pressure the resistance dropped to ~5 Ohm and then below 1 Ohm at pressurizing to 177 GPa. The step at ~180 K developed at the cooling (olive line). It became more pronounced (blue line) with time (15 hours). After pressurizing to 197 GPa at 300 K and next cooling the minimum resistance reached R=1.7 $10^{-4}$ Ohm at 144 K (inset). Corresponding resistivity $\rho$~1.7 $10^{-10}$ Ohm m ~50 times lower than for copper (at 150 K $\rho$= 70 * $10^{-10}$ Ohm m, Ref. 27).
There are notable oscillations on the R(T) pronounced at the olive curve. We observed these oscillations with period of 25-30 K in a number of runs. The resistance plots (olive and blue lines) taken in PPMS were averaged from the measurements with increasing and decreasing of temperature. The rest of the plots are measurements in optical cryostat at slow (about 5 hours) warming so the temperature was close to equilibrium.

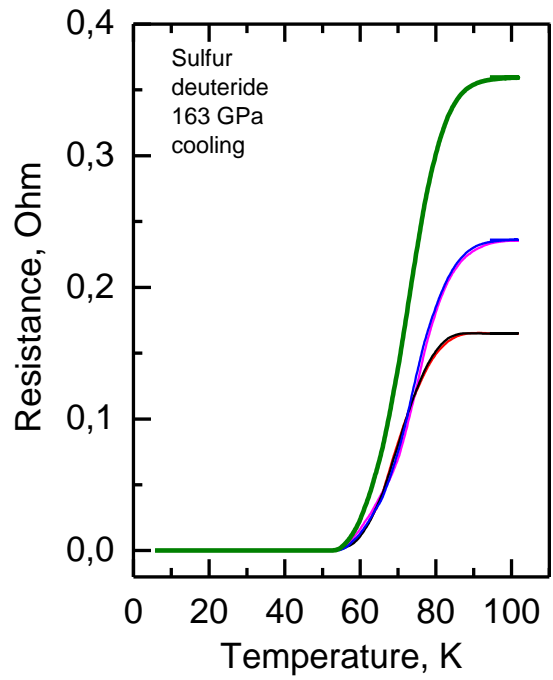

Fig. 4. Electrical measurements of sulfur deuteride at 163 GPa. The sample was pressurized to 163 GPa at 200 K. R(T) measured with current of 10 mA at decreasing of temperature. Color lines – measurements on four channels, and olive line – the resistance calculated from these data with the van der Pauw method.

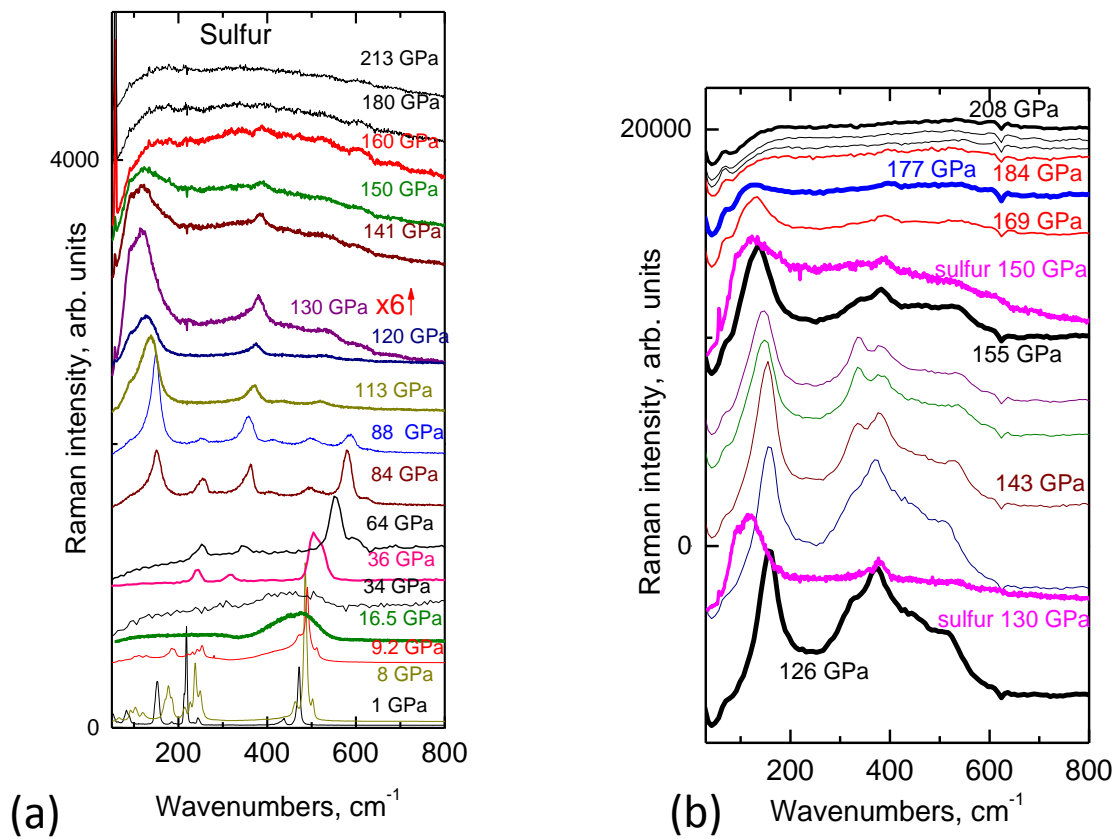

Fig. 5. Raman spectra of sulfur hydride compared with elemental sulfur. Ultralow luminescence synthetic diamond anvils allowed us to record the Raman spectra at high pressures in the metallic state.
(a) Raman spectra of elemental sulfur at different pressures measured at room temperature in the wide pressure range. At small pressures Raman signal is very strong but above ~10 GPa the intensity dramatically decreases, some reflectivity appears. Above 80 GPa a new phase appeared which persists to 150 GPa in the metallic state. At 160 GPa Raman disappears, likely because of transformation to the β-Po phase. The pressure of the transformation is in a good agreement with our four probe electrical measurements of $T_c$ (Fig. 2a). The electrical measurements in turn are good agreement with the susceptibility measurements (Gregoryanz *et al,* Phys. Rev. B **65**, 064504) but we obtained noticeable higher $T_c$s at P> 200 GPa (to be published).
(b) Raman spectra of sulfur hydride at releasing pressure from 208 GPa at room temperature. The spectra are much stronger than those from sulfur in the metallic state at high pressures. There is apparent phase transition at ≈180 GPa.

There is apparent difference in the Raman spectra of sulfur hydride (a) and sulfur (b): the peaks at ~100 cm$^{-1}$ shift each other, phase transformation are at different pressures: 160 GPa for sulfur and 180 GPa for sulfur hydride. Finally, Raman spectra of sulfur hydride are significantly stronger.